\documentclass[prb,reprint]{revtex4-1} 

\usepackage{amsmath} 
\usepackage{graphicx} 
\usepackage{subfigure}

\begin{document}

\title{Laboratory limit on the charge of photons by electric field deflection}
\author{A. Hankins}
\author{C. Rackson}
\author{W. J. Kim}
\email{kimw@seattleu.edu}
\affiliation{901 12th Ave Department of Physics, Seattle University, Seattle, WA 98122, USA}
\date{\today}

\begin{abstract}
The deflection of a laser beam traveling through a modulated electric field is measured using phase-sensitive detection to place an upper bound on photon charge. An upper limit of $10^{-14}e$ is obtained. The experiment involves a number of experimental techniques that are commonly encountered in modern precision measurements and is suitable for both advanced undergraduate and beginning graduate students in physical science as a laboratory exercise.   
\end{abstract}

\pacs{14.70.Bh}
\keywords{Suggested keywords}
          
\maketitle

\section{Introduction}
Light is described as a stream of particles called photons, which propagate at the finite speed $c=2.998\times10^8$ m/s. Unlike a stream of electrons, which are deflected under the influence of external fields, these photons, when traveling through a region of electric (or magnetic) field, have not been observed to deflect. Even more puzzling is that they are completely insensitive to the presence of the very fields that give rise to their own existence: the quanta of electromagnetic waves, though generated by accelerating electric charges, become indifferent to the same electric charges upon creation.   

Essentially, photons are elementary bosonic particles with zero mass and zero charge. The former property---photons possessing no mass---underlies the fundamental interaction between two electric charges described by Coulomb's inverse-square law. The fact that the inverse-square law holds with extreme precision, directly translates into validity of photons' having zero mass, and vice-versa. Precision tests of the inverse-square law through Gauss's law and its divergence theorem have been performed over the decades, and the best laboratory limit\cite{faller} of the mass of photons on the order of $10^{-50}$ kg was obtained by Williams, Faller, and Hill in 1971.

The latter property---photons possessing no electric charge---is a bit trickier to imagine, simply because there is no complete theory that predicts otherwise. One possible scenario is that photons are single charged particles carrying the same magnitude of electric charge; however, the fact that electromagnetic fields themselves (made up of these same photons) have never been observed to be charged suggests that the single-charge photon model might be implausible, affirming that fields and charges are inherently different. Another scenario is to have photons with multiple charges (e.g. two types of photon with opposite signs of charge or three different types including a neutral one), thereby obeying overall charge neutrality. This is essentially to conjecture a non-Abelian gauge theory in which a nontrivial multiplet of interacting bosons exists. The problem with this scenario is that the charge of the photon must be quantized in units of the elementary electric charge, but the observed photon charge is many orders of magnitude smaller than $e$. The extra degeneracy associated with the photons therefore appears to contradict the totality of data in the Standard Model, as pointed out by Okun\cite{okun1,okun2}. 

The quest to understand the ultimate nature of photon goes back to 1932 when de Broglie proposed that a photon is composed of a neutrino and an antineutrino, satisfying the spin-1, zero-mass, and zero-charge characteristics of the photon.\cite{deb} In this model, photons are composite particles---rather than elementary particles---described as a bound state formed by a pair of neutrinos obeying Fermi-Dirac statistics (i.e. photons are fermions). The neutrino theory of photons was further developed by W. A. Perkins\cite{neutrino} in 1967, but the work generated little interest and no dedicated experimental activity has been performed to fully reveal the photon's true nature. Obviously, the mere absence of a complete theoretical framework is no reason not to undertake precision tests on the fundamental properties of photons, primarily because such tests are not only crucial to our basic understanding of electromagnetic theory, but also they provide a valuable opportunity to assess how well the existing theory has been experimentally tested, particularly in a laboratory setting.\cite{rev} 

\begin{table*}[htbp]
\centering
\begin{tabular}{ @{} ccccc @{} }
\hline
Authors & Year & $q/e$ & Interaction source & Photon source\\
\hline
\hline
Grodzins et al.\cite{grod} & 1961 & $10^{-15}$ & Electric field & Decay of Fe$^{57}$ (laboratory test)\\
Stover et al.\cite{stover} & 1967 & $10^{-16}$ & Electric field & Laser (laboratory test)\\
Present work & 2012 & $10^{-14}$ & Electric field & Laser (laboratory test)\\ 
\hline
\hline
Cocconi\cite{cocconi} & 1992 & $10^{-28}$ & Magnetic field & Extragalactic radiation\\
Cocconi \& Raffelt\cite{cocconi2,raffelt} & 1994 & $10^{-28}$ & Magnetic field & Radio pulsars\\
Sivaram\cite{sivaram3}  & 1994 & $10^{-27}$ & Charge asymmetry & CMB\\
Semertzidis et al.\cite{lazarus} & 2003 & $10^{-16}$ & Magnetic field & Laser (laboratory test)\\
Kobychev and Popov\cite{popov} & 2005 & $10^{-31}$ \& $10^{-33}$ & Magnetic field & Extragalactic radiation\\
Caprini and Ferreira\cite{caprini} & 2005 & $10^{-38}$ & Charge asymmetry & CMB\\
Altschul\cite{altschul} & 2007 & $10^{-32}$ \& $10^{-46}$ & Magnetic field & Extragalactic radiation\\
Sivaram and Arun\cite{sivaram} & 2010 & $10^{-30}$ & Magnetic field & Hawking radiation from a black hole\\

\hline
\end{tabular}
\caption{Status of upper limits on the photon charge: There have been three laboratory tests to constrain the photon charge; two of them employ electric fields, one by Grodzins et al.\cite{grod} and the other by Stover et al. \cite{stover}, and the third one involves magnetic deflection \cite{lazarus}. Even the best limit among the three laboratory tests is at least ten orders of magnitude weaker than the limit obtained from charge asymmetry in the CMB data \cite{sivaram3}, which is the least stringent bound among those based on astrophysical observations.}
\end{table*} 

According to the 2012 Particle Data Group (PDG)\cite{pdg}, the best limit on the charge of a photon has been obtained by looking for Aharonov-Bohm phase differences from extragalactic radiation.\cite{altschul} The non-observation of such phase differences has placed an upper bound on the photon charge at the level of $10^{-32}e$, assuming all photons have the same charge. Several other limits based upon similar astrophysical observations have been reported since 1988. Examples include: limits derived from the analysis of the energy spreads of incoming photons from both radio pulsars \cite{cocconi2,raffelt} and extragalactic sources \cite{cocconi,popov}, and the cosmic microwave background (CMB) \cite{sivaram3,caprini} (See Table 1 for a summary of the current status on the bounds of the photon charge). In contrast, only two laboratory limits---using an electric field as a source of interaction---exist in the literature, one by Grodzins et al.\cite{grod} in 1961 and the other one by Stover et al.\cite{stover} in 1967. In fact, the photon charge limit reported by Stover and his coworkers was the by-product of an experiment originally intended at searching for an electron-proton charge inequality. Additionally, the naive procedure to place a limit by calculating a total change of the electric charge on a spheroid per unit photon has been cast into serious doubt.\cite{altschul2} From this perspective, the experiments by Grodzins et al. remains the only laboratory experiment solely dedicated to testing on the charge of a photon involving electric fields, although the work has not been published in a peer-reviewed journal. 

Here, we report an experimental test placing an upper limit on the charge of a photon using a light beam traveling through an electric field with a phase-sensitive detection technique. Our experimental arrangement is distinct from previous tests constraining limits on the photon charge for several reasons: (i) We employ electric, rather than magnetic deflection; all previous reports have exclusively employed deflection by magnetic fields, either from astrophysical data \cite{cocconi,raffelt,popov,altschul} or using an electrically tunable magnet \cite{lazarus}. (ii) Our simple experimental scheme can be readily implemented in undergraduate laboratories, introducing students to a number of experimental techniques that are commonly employed in modern precision measurements. These include: phase-sensitive detection, precision capacitance measurements, Michelson interferometry, and phase-delay measurements. (iii) With some modifications to the present setup, one could place a stronger limit on the photon charge by at least two orders of magnitude; any modest improvement to the current limit should be regarded as a valuable contribution to the field, particularly when the limit is obtained from a laboratory test in which all the relevant parameters are controlled.    

\section{Experimental apparatus}
The experimental configuration is depicted in Fig. 1. A laser beam of $\lambda=667$ nm travels through a circular parallel-plates capacitor whose diameter is $D=5.08\pm 0.01$ cm. The signal from a function generator (Agilent 33120A) is amplified by a high voltage power supply ($\Delta V_{\rm{max}}=265$ V) which then modulates the voltage across the parallel-plate. The signal from the function generator also provides a reference input for a lock-in amplifier (Stanford Research 830), and the resulting ``deflection'' signal from a position sensitive photodiode is demodulated by the lock-in. Not shown in Fig. 1 is an additional mirror employed to reflect the beam, which is made to travel 50 m down a hallway and back, making a total leveraging distance of $L=100$ m. The angular deflection is then expressed as $\Delta\theta=\Delta x/L$, where $\Delta x$ is the $A-C$ signal representing the horizontal displacement of the quadrant photodiode. 

\begin{figure}[htbp]
\centering
\includegraphics[width=1.0\columnwidth,clip]{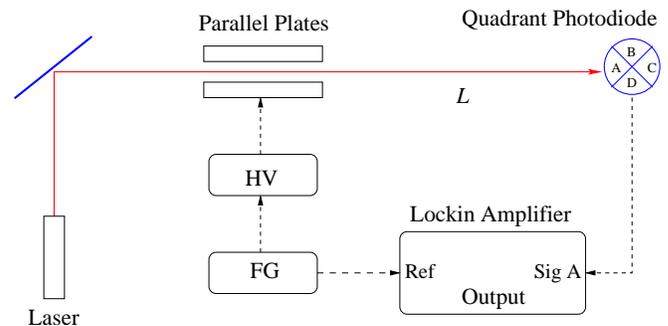}
\vspace{0.5cm}
\caption{Experimental setup: The electric field ($V/d)$ across the parallel plates is modulated with the upper plate grounded, and the subsequent light ``deflection'' is measured from the lock-in output.} 
\label{fig1}
\end{figure}

The angular resolution (i. e. minimum detectable signal $\Delta\theta_{\rm{min}}$) is determined in the following way: A piezoelectric transducer (PZT) is attached to a linear stage, which then physically moves the photo detector back and forth at a modulation frequency (i.e. it mimics deflections). The output of the lock-in amplifier is plotted as a function of the modulation voltage applied to the PZT, as shown in Fig. 2. The lock-in response is linear, and its sensitivity can be inferred from the smallest PZT voltage modulation it is able to detect, which is measured to be $\Delta V_{\rm{PZT}}^{\rm{min}}$=103 mV. Note that $\Delta V_{\rm{PZT}}$ corresponds to a certain distance that the PZT travels (i .e. $\Delta x$). 

\begin{figure}[htbp]
\centering
\includegraphics[width=1.0\columnwidth,clip]{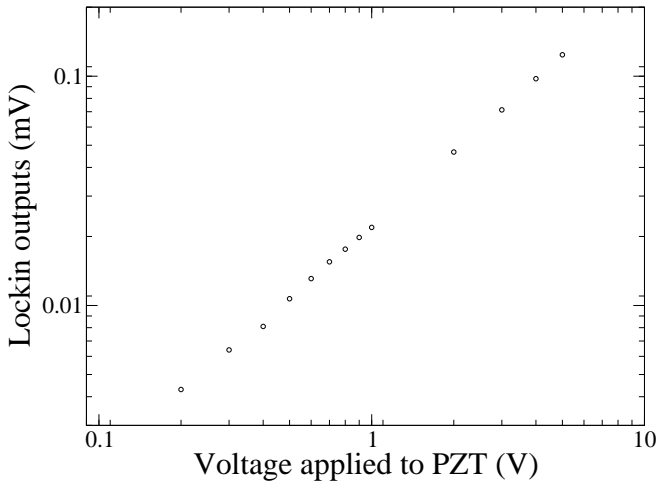}
\vspace{0.5cm}
\caption{Calibration for the lock-in amplifier: The larger the voltage applied to the PZT, the greater its displacement and hence the larger the output response from the lock-in. Care is taken not to drive the PZT with a negative voltage; an offset voltage is applied to ensure the modulation signal is always positive. Similar calibration data are also obtained at different modulation frequencies.}
\label{fig2}
\end{figure}

Next, we calibrate the voltage applied to the PZT with the actual distance traveled by it. This is achieved by implementing a Michelson's interferometer, as shown in Fig. 3. The period of interference intensity is related to $\lambda_c/2$, where $\lambda_c$ is the wavelength of a Helium-Neon laser. From this, the PZT distance-to-voltage ratio ($\Delta x/\Delta V_{\rm{PZT}}$) is measured to be $127\pm1$ nm/V. This yields a deflection sensitivity of $\Delta x_{\rm{min}}=13$ nm and thus an angular sensitivity of $\Delta\theta_{\rm{min}}=1.3\times10^{-10}$ radians. 
\begin{figure}[htbp]
\centering
\includegraphics[width=0.9\columnwidth,clip]{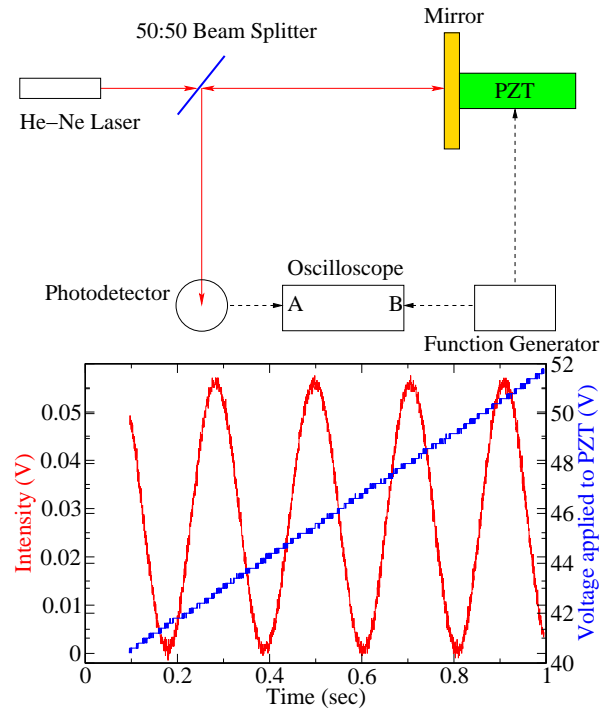}
\includegraphics[width=0.9\columnwidth,clip]{fig3b.eps}
\vspace{0.5cm}
\caption{Schematic of a Michelson's interferometer (top) and interference patterns (bottom): The intensity resulting from an interference between two light beams (i.e. one traveling directly into the photodiode and the other one traveling a longer distance and being reflected from a mirror attached to the PZT. As the PZT moves, the intensity varies from minimum (destructive) to maximum (constructive) with a visibility larger than 90\%. Periodicity of $\lambda_c/2$, where $\lambda_c=632.8$ nm enables a direct calibration of the PZT's applied voltage to the actual position displacement.}
\label{fig3}
\end{figure}

Another important parameter to be characterized is the absolute separation distance $d$ of the parallel-plates. This is most effectively achieved by measuring the capacitance at different gap separations, which is expected to obey the $1/d$ power law for parallel-plate geometry through $C=\epsilon_0A/d$, where $\epsilon_0$ is the permittivity constant in vacuum and $A=\pi (D/2)^2$ is the plate area. 

In Fig. 4, a circuit diagram of a simple relaxation oscillator is presented along with the measurement data in which the oscillation period is plotted against the separation distance. At the heart of the relaxation oscillator is an op-amp (OP27) which acts as a comparator; its oscillation period is proportional to the unknown capacitance $C_{\rm{pp}}$ and is expressed as $T(d)=2RC(d)\ln3$, where $R=1$ M$\Omega$ and $C$ is the total capacitance including the 47 pF capacitor. 

The purpose of the capacitance measurements is two-fold: First, it enables the extraction of a point of contact $d_0$ from which the absolute distance is determined from a relative displacement $d_r$. Second, more importantly, it identifies a distance range in which the parallel-plate formula is validated. At a short separation distance (i.e. $d<0.25$ mm), the data tends to suffer from a severe deviation from the expected $1/d$ power law. Because of finite-parallelism, our measurements have been applied to a distance range from 1.5 mm to 0.25 mm, and a separation distance of $d$=1.3 mm is maintained throughout the experiment. The relaxation oscillator is quite useful in its own right and could be applied in another independent project run by students. For instance,  various power laws governing the capacitance versus distance in different geometries could be studied, as they have raised a series of discussions in recent Casimir force measurements.\cite{ana,cas}   

\begin{figure}[htbp]
\centering
\includegraphics[width=0.9\columnwidth,clip]{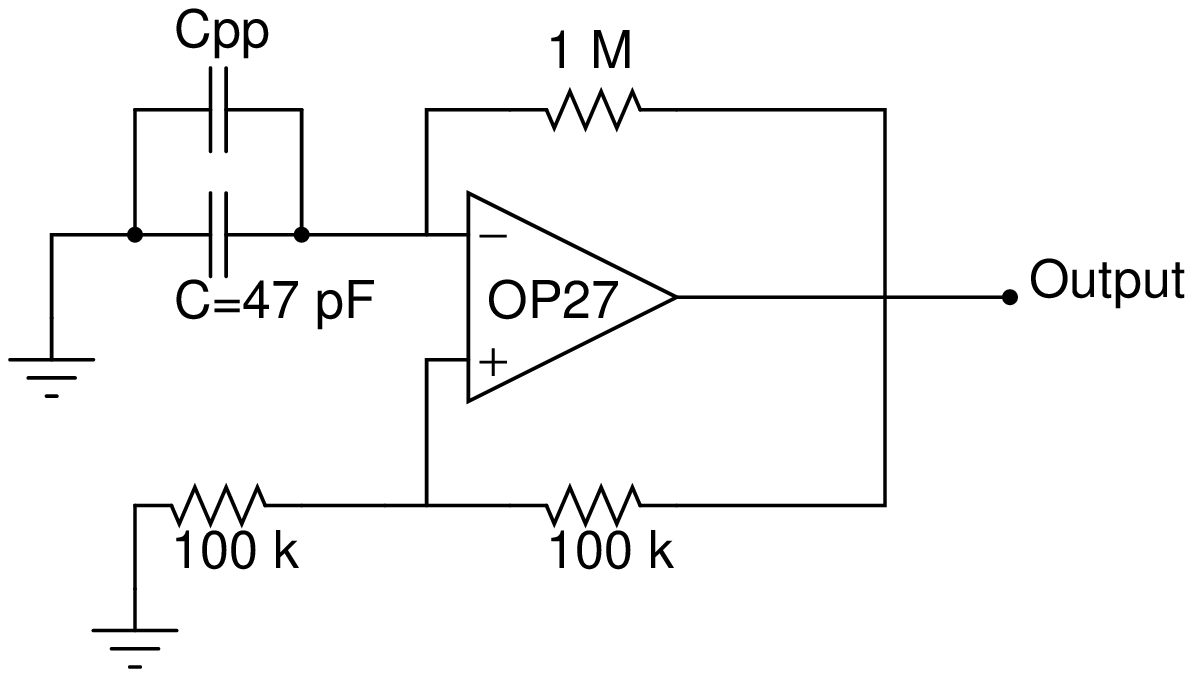}
\includegraphics[width=0.9\columnwidth,clip]{fig4b.eps}
\vspace{0.5cm}
\caption{Circuit diagram of a relaxation oscialltor (top) and period measurements (bottom): The period of the relaxation oscillator (i.e. output) is proportional to the external capacitance. A motorized actuator moves one of the plates, thereby changing their separation distance. The absolute distance is not known {\textit {a priori}}, and measured data are fit to a function $T=T_0+\beta/(d_0-d_r)$, where $T_0$ represents an offset capacitance (or $C_0$ in the limit of $d\rightarrow\infty$). $d_0$ is a point of contact to be determined from the fit and $d_r$ is the relative position of the actuator. Finally, $\beta$ is the period-to-capacitance conversion factor and contains the information regarding the effective size of the plate as well as the permittivity constant in between. The absolute distance characterization is an essential part of calibration procedures in precision force measurements\cite{cas}.}
\label{fig4}
\end{figure}

\section{Results and analysis}
Based upon elementary considerations of a relativistic, massless particle whose charge $q$ experiences an electrostatic force inside a parallel-plate, the photon's deviation from its straight path is given by 
\begin{equation}
\theta=qVl/Ed, 
\end{equation}
where $E=hc/\lambda$ is the photon energy with $h$, Planck's constant. $l$ is the effective interaction length corresponding to the diameter $D$ of the plates employed in our experiment. The above expression can be also related to the previous experiments constraining the charge of the photon by magnetic deflection in which the angular deflection on the detector is expressed as:
\begin{equation}
\Delta\theta=\frac{ql}{Ed}\left\{\Delta V+ V\frac{\Delta E}{E}\right\}.
\end{equation}
The first term in the brackets is the feature of an experiment in which the field strength is modulated. If no modulation on the field strength is applied (i.e. $\Delta V=0$), one is left with the second term which represents the angular spread of the incoming photons, within a bandwidth $\Delta E$. Assuming no modulation and replacing the field magnitude $V/d\rightarrow Bc$, the upper limit on the charge of the photon becomes 
\begin{equation}
\frac{q}{e}<\frac{E^2\Delta \theta}{\Delta EBcle}.
\end{equation}
Aside from the absence of a factor of two arising from a binomial expansion, Eq. (3) exactly matches the derived results\cite{cocconi,altschul} for the case of the deflection of a photon that has traveled a distance $L$ through the intergalactic magnetic field $B$. Note that less energetic photons and larger bandwidths are needed to constrain a stronger limit, as was applied in the previous experiments. 

It is clear from Eq. (2) that unless $\Delta E\gg E$, a greater sensitivity is achieved by modulating the strength of an external field, leading to the upper limit on the photon charge: 
\begin{equation}
\frac{q}{e}<\frac{E\Delta \theta}{(\Delta V/d)le}.
\end{equation}
Replacing the modulation of the electric field with that of a magnetic field $\Delta V/d\rightarrow\Delta Bc$, one also retrieves the upper limit of the photon charge obtained by Semertzidis et al.\cite{lazarus}, which was reported to be $8.5\times 10^{-17}e$. It should be noted that the charge limit obtained in all experiments involving magnetic deflection assume the cyclotron trajectory of the photon in a transverse magnetic field. There, the linear momentum of the photon is conserved in a Larmor radius given by $R=p/qB$, where $p=hf/c$ is the relativistic momentum of the photon, $f$ is the oscillation frequency of the photon; because magnetic forces do no work, the photon energy is independent of its mass and its momentum remains constant, given by $p=E/c$. 

The situation is somewhat different in the case of the electric field deflection. Because the linear momentum associated with the direction of the applied electric field (i.e. the direction in which the photon is deflected) is not conserved when it travels through the electric field, the photon gains electric potential energy. In fact, the gain of energy of photons crossing an electric field can be directly measured by determining the absorption condition of resonance in the decay of, for example, Fe$^{57}$ based upon the M{\"o}ssbauer effect\cite{grod}---the underlying technique in the earliest experiment aimed to measure the charge of a photon.   

We now place an upper limit on the charge of the photon in our experimental situation. Using the angular resolution $\Delta\theta_{\rm{min}}=1.3\times10^{-10}$ radians and the separation distance $d=$1.3 mm, we attain
\begin{equation}
q=\frac{hc\Delta\theta}{(\Delta V/d)l\lambda}<2.3\times10^{-14}e,   
\end{equation}
which is remarkable given the relatively straightforward approach adopted in an undergraduate laboratory. This is only two orders of magnitude weaker than the most stringent limit obtained from a laboratory test involving magnetic deflection\cite{lazarus}. Clearly, one can easily notice an advantage in employing magnetic deflection with the speed of light $c$ as a multiplication factor in the field strength (i. e $\Delta V/d\rightarrow\Delta Bc$). However, a parallel-plate is easy to implement and requires only a modest amplification of the applied voltage for placing a strong limit. In principle, it is possible to improve the present limit by at least two orders of magnitude; this is achieved in part by increasing both the modulation voltage to about 2 kV (before the electric field breakdown in air) and the interaction length to about $l=50$ cm. 

One can argue that Rayleigh scattering between photons and air molecules should be taken into account. Note that our phase-sensitive detection is only sensitive to the modulation frequency, and no ``deflection'' was registered at the output of the lock-in amplifier. This suggests that in the presence of any systematic effects leading to spurious ``deflections'' the quoted sensitivity corresponds to the maximum charge of the photon (i.e. the actual charge of the photon could be much smaller). 

We have also independently measured the speed of light of the employed beam to investigate any possible effects caused by the dielectric permittivity in air. This was performed by modulating the intensity of the light beam at tens of MHz and measuring the phase difference originating from a time delay over a certain distance (see Fig. 5). 
\begin{figure}[htbp]
\centering
\includegraphics[width=0.9\columnwidth,clip]{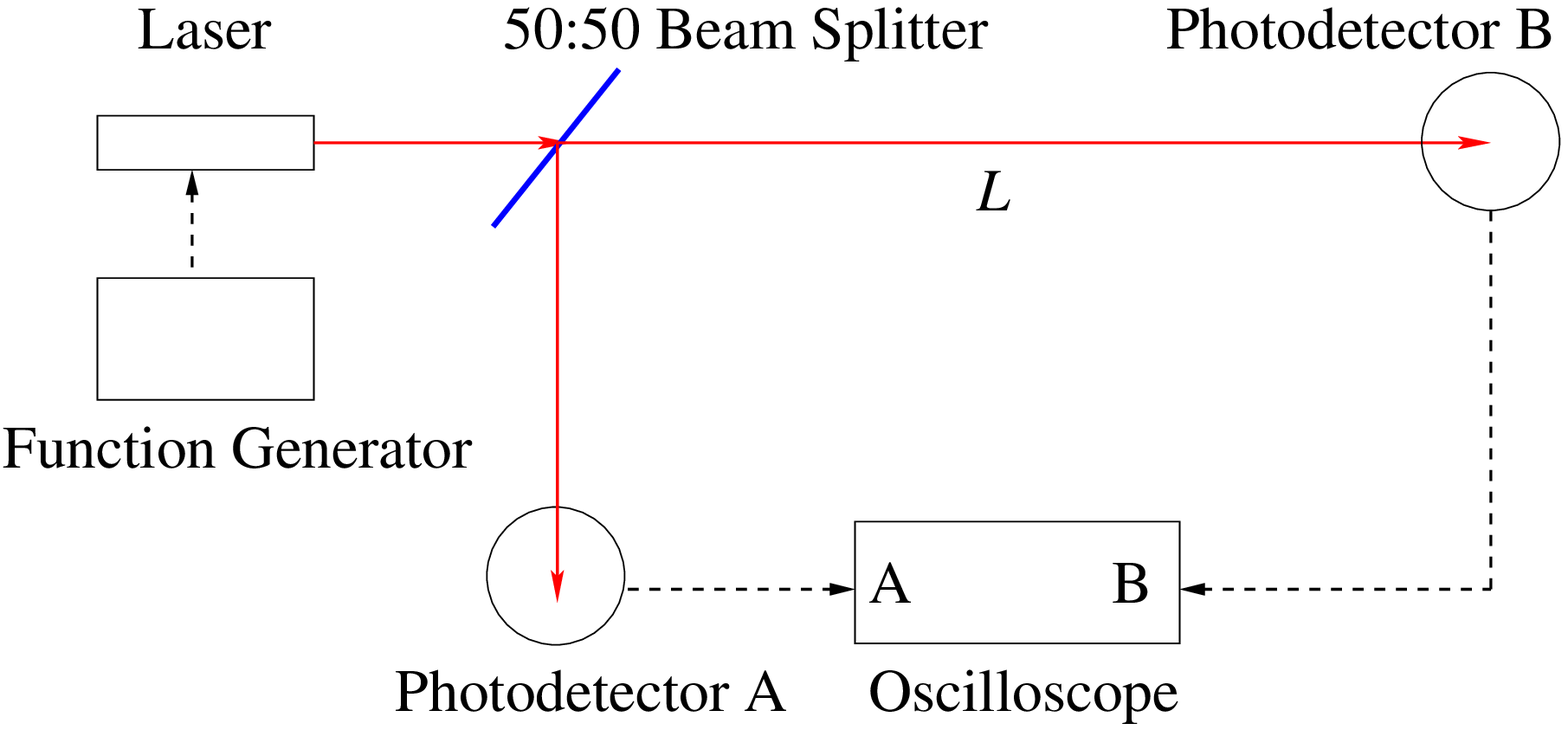}
\includegraphics[width=0.9\columnwidth,clip]{fig5b.eps}
\vspace{0.5cm}
\caption{Schematic diagram to determine the speed of light (top) and phase change (time delay) versus distance (bottom): The laser used in our experiment is modulated at 15 MHz and is split into two different paths. Delay times associated with different travel lengths are then measured in terms of phase changes between two photodetectors \cite{speedoflight}. A linear fit to the data leads to $c=(2.999\pm0.006)\times10^8$ m/s. Modulation frequencies ranging from 1 MHz to 15 MHz are used, leading to similar results.}
\label{fig5}
\end{figure}
The measured value agrees with the accepted, and no difference was observed from the speed of light in vacuum within our experimental uncertainties.

\section{Conclusion}
We have reported the result of an experimental test placing a limit on the photon charge by deflection in an electric field. An upper bound on the order of $10^{-14}e$ has been obtained \cite{note}. Our relatively straightforward arrangement can be easily implemented in an undergraduate laboratory in which students gain exposure to a variety of experimental techniques. The entire experiment--- including all the calibrations involving a Michelson's interferometer, the construction of a relaxation oscillator, and the speed of light measurements---could be completed in a period of several weeks by one or two advanced undergraduate students. Furthermore, the present limit is only two orders of magnitude away from the best laboratory limit reported to date. With some modification to the present setup, one could achieve a tighter upper bound, thereby providing a valuable contribution to ongoing searches for the electric charge of the photon.

\begin{acknowledgments}
We thank Roberto Onofrio, Paul Fontana, and Michael Morgan for providing valuable comments on our manuscript, and Jeff Wilhite for machining various parts for our experiment. This work is supported by the M. J. Murdock Charitable Trust and by Research Corporation through Single Investigator Cottrell College Science Award (SI-CCSA). A. Hankins acknowledges the research support made possible by Pat and Mary Welch.
\end{acknowledgments}

\end{document}